\documentclass[reprint,twocolumn]{revtex4-1}
\usepackage{amsfonts}
\usepackage[tbtags]{amsmath}
\usepackage{amssymb}
\usepackage{amsthm}
\usepackage{color}
\usepackage{charter}
\usepackage{graphicx}
\usepackage{lipsum}

\begin{document}

\title{Conventional and unconventional photon blockade effects in an atom-cavity system}
\author{Xinyun Liang}
\author{Zhenglu Duan}
\email{duanzhenglu@jxnu.edu.cn}
\author{Qin Guo}
\author{Cunjin Liu}
\author{Shengguo Guan}
\author{Yi Ren}

\begin{abstract}
A two-level system interacting with a cavity field is an important model for
investigating the photon blockade (PB) effect. Most work on this topic has
been based on the assumption that the atomic transition frequency is
resonant with the fundamental mode frequency of the cavity. We relax this
constraint and reexamine PB in a more general atom--cavity system with
arbitrary atomic and cavity detunings from a driving field. The results show
that when the signs of the atomic and cavity detunings are the same, PB
occurs only in the strong-coupling regime, but for opposite signs of the
atomic and cavity detunings, strong photon antibunching is observed in both
the weak- and strong-coupling regimes and a better PB effect is achieved
compared with the case when the signs are the same. More interestingly, we
find that this PB arises from quantum interference for both weak and strong
nonlinearities. These results deepen our understanding of the underlying
mechanism of PB and may be help in the construction of single-photon sources
with higher purity and better flexibility using atom--cavity systems.
\end{abstract}

\maketitle

\affiliation{College of Physics, Communication and Electronics, Jiangxi Normal University,
Nanchang, 330022, China }

\affiliation{College of Physics, Communication and Electronics, Jiangxi Normal University,
Nanchang, 330022, China }

\affiliation{College of Physics, Communication and Electronics, Jiangxi Normal University,
Nanchang, 330022, China }

\affiliation{College of Physics, Communication and Electronics, Jiangxi Normal University,
Nanchang, 330022, China }

\affiliation{College of Physics, Communication and Electronics, Jiangxi Normal University,
Nanchang, 330022, China }

\affiliation{College of Physics, Communication and Electronics, Jiangxi Normal University,
Nanchang, 330022, China }

\section{Introduction}

Single-photon sources play an important role in quantum optics and quantum
information processing \cite{1,2,3,4,5}. The quantum statistics of a
single-photon source corresponds to a sub-Poissonian distribution, which can
be realized via the photon blockade (PB) effect \cite{6,60,7}.

In the PB effect, a photon can impede the transmission of later photons. In
a strongly nonlinear quantum optics system, the strong nonlinear interaction
between photons can lead to a quantum anharmonic ladder in the energy
spectrum. If the nonlinear energy shift in the two-photon state is greater
than the loss of the cavity, population of the two-photon state may be
substantially suppressed such that only one photon is allowed in the system
at a given time: this is called conventional photon blockade (CPB) \cite{7}.
The strong nonlinearity of the system can arise from a single two-level
emitter strongly coupled to a optical cavity mode or from a three-level
emitter coupled to a cavity operating near the electromagnetically induced
transparency (EIT) window \cite{EIT}. CPB was first observed experimentally
with a trapped atom in an optical cavity system \cite{8}. Besides
atom--cavity systems \cite{9}, other quantum systems, such as
quantum-dot--cavity systems \cite{QD,11,QDsystem}, superconducting qubit
systems \cite{12,13,14}, and optomechanical systems \cite{15,16,17}, have
been used to explore CPB and serve as single sources.

In contrast, in a weakly nonlinear quantum optics system, there is another
type of PB, known as interference-based PB or unconventional PB (UCPB),
which has been studied in Refs.~\cite{18,19,20,21,22,23,24}. The physical
mechanism underlying UCPB is that there are two or more different quantum
paths for a one-photon state transiting to a two-photon state, which may
lead to quantum destructive interference on the two-photon state and result
in strong photon antibunching. A number of quantum systems have been found
to exhibit UCPB, such as two coupled optical cavities with weak nonlinearity
\cite{19} and a linear optical cavity with parametric plus coherent driving
fields \cite{PCD}. Similarly, phonon blockade based on strong nonlinearity
or a quantum interference mechanism has also been proposed and indeed
observed \cite{phonon1,phonon2}.

To date, most studies of PB in quantum two-level emitter--cavity systems
have made the assumption that the atomic detuning is equal to the cavity
detuning. However, it is experimentally difficult to fabricate a cavity with
fundamental frequency exactly resonant with the transition frequency of an
atom. Additionally, this constraint on the atomic and cavity detunings may
reduce flexibility and limit the application of single-photon sources based
on PB in atom--cavity systems. Hence, some works about photon blockade turn
to the detuned atom-cavity system and found the detuning between atom and
cavity seriously affect the photonic statistics of the system \cite%
{DPB0,DPB1,DPB2,DPB3}. In the present study, we also relax the constraint on
atomic detuning and cavity detuning and reexamine the PB effect with the
hope of discovering further interesting and important features of PB in
two-level atom--cavity systems. Most work on PB has focused on strongly or
weakly nonlinear regimes. To give a general picture of PB, we present in
this work a unified theoretical analysis of PB in a regime involving both
strong and weak coupling. This may lead us to reconsider PB arising from
nonlinear photon interactions and from quantum interference, and may provide
further understanding of the physical mechanisms underlying PB.

The remainder of this paper is organized as follows. In Sec.~\ref{secII}, we
present a physical model of the atom--cavity system and analytically study
PB, relaxing the constraint on the atomic and cavity detunings. The optimal
parameter relations for strong photon antibunching are given in the strong-
and weak-coupling regimes based on the dressed state framework and the
wavefunction method, respectively. Furthermore, general considerations about
PB in the strong- and weak-coupling regimes are presented and are used to
comprehensively illustrate the physical mechanism behind PB. In Sec.~\ref%
{secIII}, we present a numerical study of photon statistical properties via
the master equation and compare the results with the analytical results.
Finally, we conclude the work in Sec.~\ref{secIV}.

\section{Theoretical model}

\label{secII}

\begin{figure}[tbp]
\centering \includegraphics[width=1\columnwidth]{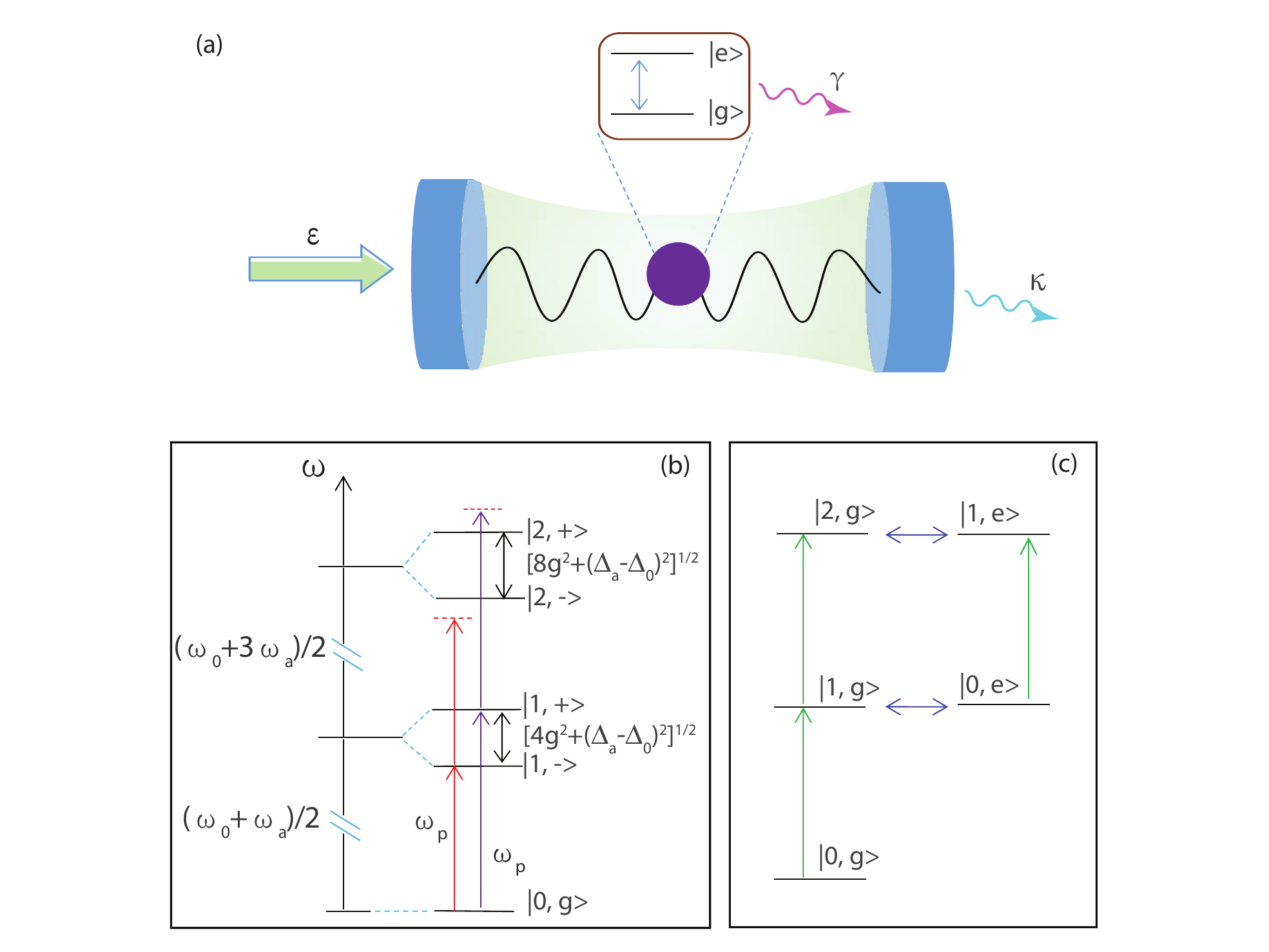}
\caption{(Color online) (a) Scheme for photon blockade of a two-level atom
coupled with a cavity. (b) Energy level diagram of dressed states in a
coupled quantum-dot--cavity system. (c) Transition paths for different
photon states.}
\label{fig1}
\end{figure}

We consider a model consisting of a single two-level emitter (with
transition frequency $\omega _{0}$) coupling to an optical cavity (with
fundamental mode frequency $\omega _{a}$), driven by a weak laser, as shown
schematically in Fig.~\ref{fig1}(a). The model discussed here is fairly
general and can be realized in a variety of quantum emitter cavity systems,
such as quantum-dot--cavity systems and superconducting qubit--cavity
systems. In a reference frame rotating with the laser frequency $\omega _{p}$%
, the interaction Hamiltonian of the atom--cavity system is given by (taking
$\hbar =1$)
\begin{equation}
\begin{split}
H_{I} ={}&\Delta _{0}\sigma _{+}\sigma _{-}+\Delta _{a}a^{\dagger
}a+g(\sigma _{+}a+\sigma _{-}a^{\dagger }) \\
&+\varepsilon (a+a^{\dagger }),  \label{H0}
\end{split}%
\end{equation}
where $\sigma _{+}$ and $\sigma _{-}$ are respectively the raising and
lowering operators of the atom, and $a^{\dagger }$ and $a$ are respectively
the creation and annihilation operators of the optical cavity mode. $g$ is
the atom--light coupling strength, $\varepsilon $ describes the driving
strength of the driving field. $\Delta _{0}=\omega _{0}-\omega _{p}$ and $%
\Delta _{a}= \omega _{a}-\omega _{p}$ are respectively the atomic detuning
and cavity detuning with respect to the driving field. The Hamiltonian (\ref%
{H0}) provides the basis for the following study.

Most current work on PB in atom--cavity systems focuses on the special case $%
\Delta _{0}=\Delta _{a}$. However, PB obtained under this condition may be
limited in application. To illustrate the hidden peculiar features of PB
under the condition $\Delta _{0}=\Delta _{a}$, we assume that the atomic and
cavity detunings can take arbitrary individual values, and we then perform a
comprehensive analysis of the PB mechanism in the absence of this condition;
i.e., we allow $\Delta _{0}\neq \Delta _{a}$.

\subsection{PB induced by strong nonlinear interaction between photons}

\label{secII-A}

To study the physical mechanism underlying PB induced by strong nonlinear
interaction between photons in an atom--cavity system, we begin by
discussing the energy levels of the system. For an atom--cavity system, from
the expression for the free Hamiltonian $H_{0}=\omega _{0}\sigma _{+}\sigma
_{-}+\omega _{a}a^{\dagger }a$, we can see that it causes transitions only
between the bare states $|n,g\rangle $ and $|n-1,e\rangle $, with respective
eigenvalues $n\omega _{a}$ and $\omega _{0}+\omega _{a}(n-1)$. The bare
states $|n,g\rangle $ and $|n-1,e\rangle $ are degenerate when $\omega
_{a}=\omega _{0}$. Here, $n$ ($n=0,1,2,\dots $) is the number of photon
excitation, and $|g\rangle $ and $|e\rangle $ identify the ground and
excited states, respectively, of the atom. Using these bare states as base
vectors, the total Hamiltonian of the system without involvement of the
driving field can be described in matrix form in the subspace spanned by the
basis vectors $|n,g\rangle $ and $|n-1,e\rangle $:
\begin{equation}
H=%
\begin{pmatrix}
n\omega _{a} & g\sqrt{n} \\[3pt]
g\sqrt{n} & \omega _{0}+(n-1)\omega _{a}%
\end{pmatrix}%
.  \label{Hm}
\end{equation}%
For a given $n$, the energy eigenvalues of the total Hamiltonian can be
expressed as
\begin{equation}
E_{n,\pm }=\frac{(2n-1)\omega _{a}+\omega _{0}\pm \sqrt{4ng^{2}+(\Delta
_{a}-\Delta _{0})^{2}}}{2},  \label{En}
\end{equation}%
which corresponds to the eigenstate (in the dressed state basis)
\begin{equation}
\begin{split}
|n,\pm \rangle ={}& C_{n,\pm }\big\{\lbrack \ E_{n,\pm }+(1-n)\omega
_{a}-\omega _{0}]|n,g\rangle  \\
& +\sqrt{n}\,g|n-1,e\rangle \big \},
\end{split}%
\end{equation}%
where $C_{n,\pm }=1/\sqrt{[E_{n,\pm }+(1-n)\omega _{a}-\omega
_{0}]^{2}+ng^{2}}$ is the normalized coefficient of the dressed state $%
|n,\pm \rangle $. In fact, energy eigenvalues (\ref{En}) is very similar
with the one in \cite{DPB1,DPB2}. In the following, we can set the largest
photon excitation number as $n=2$, because the higher photon excitation
states have very low populations when PB occurs. Therefore, we can discuss
PB involving only transitions between the dressed states $|0,g\rangle $, $%
|1,\pm \rangle $, and $|2,\pm \rangle $.

We first consider the simple case in which the atomic detuning is equal to
the cavity detuning; i.e., $\Delta _{0}=\Delta _{a}$. It has been shown that
if $\Delta _{0}=\pm g$, then the excitation $\vert 0,g \rangle
\leftrightarrow \vert 1,\pm \rangle $ will be resonant and $\vert 1,\pm
\rangle \leftrightarrow \vert 2,\pm \rangle $ far from resonant, since $( 2-%
\sqrt{2}) g\gg (\kappa ,\gamma )$. In this case, the two-photon state $\vert
2,\pm \rangle $ cannot be populated, and only the one-photon state $\vert
1,\pm \rangle $ and the vacuum state $\vert 0,g \rangle $\ exist, which is
referred as CPB \cite{optimal}. In this situation, the energy level
transition is as shown in Ref.~\cite{phonon1}, and the PB comes mainly from
the nonuniform energy level spacing induced by the nonlinear interaction
between photons.

In the following, we shall relax the constraint equating $\Delta _{0}$ and $%
\Delta _{a}$, and then analyze the transition of the system energy levels so
that we can reformulate the PB mechanism in terms of a new energy level
diagram when $\Delta _{0}\neq \Delta _{a}$, as shown in Fig.~\ref{fig1}(b).
If the driving field resonantly excites the transition $\vert 0,g \rangle
\leftrightarrow \vert 1,\pm \rangle $, then the frequency of the driving
field should satisfy $\omega _{p}=E_{1,\pm }$; i.e., in combination with
Eq.~(\ref{En}), we can rewrite the resonance condition as
\begin{equation}
\omega _{p}=\left[ \omega _{a}+\omega _{0}\pm \sqrt{4g^{2}+(\Delta
_{a}-\Delta _{0})^{2}}\right]\!\Big /2,  \label{5}
\end{equation}%
which can be reduced to
\begin{equation}
g^{2}=\Delta _{a}\Delta _{0}.  \label{6}
\end{equation}
Obviously, the atomic detuning and the cavity detuning must have the same
sign; i.e., both must be blue or both red. Meanwhile, the transition between
the dressed states $\vert 1,\pm \rangle $ and $\vert 2,\pm \rangle $ is far
from resonance under driving with a large detuning $[ \mp( 3\Delta
_{a}+\Delta _{0}) - ( 6\Delta _{a}\Delta _{0}+\Delta_{a}^{2}+\Delta_{0}^{2})
^{1/2}] / 2$. Hence, there is at most one photon in the cavity at any given
time. This result shows that even though $\Delta _{a}\neq \Delta _{0}$,
there is still PB, and thus the conventional condition $\Delta _{a}=\Delta
_{0}$\ for PB with strong atom--light coupling is relaxed. Here we call Eq.~(%
\ref{6}) the optimal condition for strong PB when $\Delta _{a}\neq \Delta
_{0}$. It should be noted that if $\Delta _{a}=\Delta _{0}$, then the
optimal condition for PB in this work will reduce to the conventional one, $%
\Delta _{0}=\pm g$. Furthermore, from the above discussion and the energy
levels in Fig.~\ref{fig1}(b), we can conclude that the physical mechanism
underlying the PB considered in this subsection is nonuniform energy level
spacing between different excitation states induced by the nonlinear
interaction between photons when $\Delta _{a}\Delta _{0}>0$.

\subsection{PB based on quantum interference}

\label{secII-B}

As well as strongly nonlinear interaction between photons, quantum
interference is also a way in which PB can be realized, and a form of the
quantum interference effect is shown in Fig.~\ref{fig1}(c). The interference
can be described by two paths: the direct excitation $\vert 1,g \rangle
\rightarrow \vert 2,g \rangle $ and the tunnel-coupling-mediated transition $%
\vert 1,g \rangle \rightarrow \vert 0,e \rangle \rightarrow \vert 1,e
\rangle \rightarrow \vert 2,g \rangle $. In this subsection, we will explore
the physical mechanism underlying the generation of PB by quantum
interference when the constraint on atomic detuning and cavity detuning is
relaxed.

Following the approach presented in Ref.~\cite{19}, in the weak-driving
limit, we expand the wavefunction of the system in terms of the bare states
and just up to the two-photon excitation state:
\begin{equation}
\begin{split}
\vert \Psi \rangle ={}&C_{0,g} \vert 0,g \rangle +C_{1,g} \vert 1,g \rangle
+C_{0,e} \vert 0,e \rangle \\
&+C_{2,g} \vert 2,g \rangle +C_{1,e} \vert 1,e \rangle ,
\label{wavefunction}
\end{split}%
\end{equation}
where the coefficients $\vert C_{n,g} \vert ^{2}$ and $\vert C_{n,e} \vert
^{2}$ denote the probabilities of the system in the states $\vert n,g
\rangle $ and $\vert n,e \rangle $, respectively. Usually, we employ the
equal-time second-order correlation function $g^{( 2) }( 0) $ to measure the
quantum statistics of the optical field in the cavity when the system is in
a steady state \cite{28}. If $g^{( 2) }( 0) <1$, then the photons are
antibunched, which means that they do not like to share the same cavity. In
the steady state (i.e., $t\rightarrow +\infty $), the equal-time
second-order correlation function of the optical field can be expressed as
\begin{equation}
g^{(2)}(0)=\frac{ \langle \Psi \vert \langle a^{\dagger }a^{\dagger }aa
\rangle \vert \Psi \rangle _{s}}{( \langle \Psi \vert a^{\dagger }a \vert
\Psi \rangle _{s}) ^{2}}\simeq \frac{2 \vert C_{2,g} \vert ^{2}}{ \vert
C_{1,g} \vert ^{4}},  \label{g2}
\end{equation}%
where $\vert \Psi \rangle _{s}$ is the steady-state wavefunction of the
system. To obtain the steady-state wavefunction, we start from the Schr\"{o}%
dinger equation $i\partial _{t} \vert \Psi \rangle =H_\mathrm{eff} \vert
\Psi \rangle $\cite{21}, where $H_\mathrm{eff}$ is the effective
non-Hermitian Hamiltonian
\begin{equation}
H_\mathrm{eff}=H_{I}-i\frac{\gamma }{2}\sigma _{+}\sigma_{-} -i\frac{k}{2}%
a^{\dagger }a.  \label{nonHH}
\end{equation}%
We substitute the wavefunction (\ref{wavefunction}) and the non-Hermitian
Hamiltonian (\ref{nonHH}) into the Schr\"{o}dinger equation and set $\{ \dot{%
C} _{n,g},\dot{C}_{n,e} \} =0$, to obtain a set of equations for the
coefficients of the wavefunction:
\begin{align}
C_{1,g}g+C_{0,e}\Delta _{0}^{\prime} &=0,  \label{10} \\
C_{0,g}\varepsilon +C_{0,e}g+C_{1,g}\Delta _{a}^{\prime} &=0,  \label{11} \\
\sqrt{2}\,\varepsilon C_{1,g}+\sqrt{2}\,gC_{1,e}+2\Delta
_{a}^{\prime}C_{2,g} &=0,  \label{12} \\
\sqrt{2}\,gC_{2,g}+C_{1,e}( \Delta _{0}^{\prime}+\Delta _{a}^{\prime})
+\varepsilon C_{0,e} &=0,  \label{13}
\end{align}
with $\Delta _{0}^{\prime}=\Delta _{0}-i\gamma /2$ and $\Delta
_{a}^{\prime}=\Delta _{a}-ik/2$. The expressions (\ref{10})--(\ref{13}) also
directly illustrate the energy levels and the links between the steady
states $\vert n,g \rangle $ (or $\vert n,e \rangle $) under the weak driving
in Fig.~\ref{fig1}(c). After straightforward calculations, we obtain the
coefficients $C_{1,g}$\ and $C_{2,g}$ as follows:
\begin{align}
C_{1,g} &=\frac{\varepsilon \Delta _{0}^{\prime}}{g^{2}-\Delta
_{a}^{\prime}\Delta _{0}^{\prime}},  \label{C1} \\[6pt]
C_{2,g} &=\frac{\varepsilon ^{2}[(\Delta _{a}^{\prime}+\Delta
_{0}^{\prime})\Delta _{0}^{\prime}+g^{2}]}{\sqrt{2}\,(\Delta
_{a}^{\prime}\Delta _{0}^{\prime}-g^{2})[(\Delta _{a}^{^{\prime }}+\Delta
_{0}^{\prime})\Delta _{a}^{\prime}-g^{2}]}.  \label{C2}
\end{align}

According to the expression for the equal-time second-order correlation
function, if $C_{2,g}=0$, then $g^{(2)}(0)=0$. In this situation, the
photons will exhibit strong antibunching, and UCPB will occur. Therefore,
based on Eq.~(\ref{C2}), the optimal condition for the strong antibunching
effect is
\begin{equation}
\begin{split}
0 ={}&4( \Delta _{a}+\Delta _{0}) \Delta _{0}+4g^{2}-( \gamma +k) \gamma \\
&-2i[ \Delta _{a}\gamma +( k+2\gamma ) \Delta _{0}] .  \label{optimalC}
\end{split}%
\end{equation}
For Eq.~(\ref{optimalC}) to be valid, when $\Delta _{0}\neq 0$ and $\Delta
_{a}\neq $ $0$, the signs of the cavity and atomic detunings must be
opposite; i.e., $\Delta _{0}\Delta _{a}<0$.

The expression (\ref{optimalC}) is a general optimal condition for PB to be
induced by quantum interference, with the basic requirement that the atomic
detuning and cavity detuning have opposite signs. The physical mechanism in
this case is completely different from that discussed in Sec.~\ref{secII-A},
where the atomic and cavity detunings have the same sign. From Eq.~(\ref%
{optimalC}), one can see that if $\Delta _{0}=\Delta _{a}$, then both atomic
detuning and cavity detuning should be zero, i.e., $\Delta _{0}=\Delta
_{a}=0 $, for the optimal condition (\ref{optimalC}) to be valid. In this
situation, the real part of the optimal condition (\ref{optimalC}) will
reduce to $g= \sqrt{( \gamma +k) \gamma }/2$. These results are the same as
those obtained in Ref.~\cite{phonon1}. Of course, the optimal condition (\ref%
{optimalC}) can provide us with greater flexibility and more possibilities
to obtain strongly antibunching photons.

We rewrite the optimal parameter relation (\ref{optimalC}) as follows:
\begin{align}
\Delta _{a} &=-\left( \frac{k}{\gamma }+2\right) \!\Delta _{0},  \label{IC}
\\[3pt]
\Delta _{0} &=\pm \sqrt{\frac{g^{2}\gamma }{k+\gamma }-\frac{\gamma ^{2}}{4}
}\, .  \label{RC}
\end{align}
It can be seen that if the coupling strength $g=\sqrt{( k+\gamma ) \gamma }%
/2 $, there is only one set of optimal parameters ($\Delta _{0}=\Delta _{a}=0
$) for strong antibunching. Otherwise, there are two sets of symmetric
optimal parameters for strong antibunching.

\subsection{A unified analysis on PB}

In Secs.~\ref{secII-A} and \ref{secII-B}, we have used energy level
transition and wavefunction methods to demonstrate the existence of PB and
analyze the underlying physical mechanisms in an atom--cavity system while
relaxing the constraint on the atomic and cavity detunings. The results show
that when the signs of the atomic detuning and the cavity detuning are the
same, strongly nonlinear interaction between photons can lead to PB even if $%
\Delta _{0}\neq \Delta _{a}$, with the corresponding optimal parameter
condition being given by Eq.~(\ref{6}). When the atomic and cavity detunings
have opposite signs, and PB arises from quantum interference, the
corresponding optimal parameter conditions for strongly antibunching photons
are expressed by Eq.~(\ref{optimalC}).

Here, we will present a comprehensive discussion of the results obtained in
Secs.~\ref{secII-A} and \ref{secII-B} with the aim of understanding the
underlying physics in greater detail. From the expression $g^{(2)}(0)\simeq
2 \vert C_{2,g} \vert ^{2}/ \vert C_{1,g} \vert ^{4}$, we can see that
necessary conditions for strong photon antibunching are that $C_{2,g}$ be
sufficiently small or $C_{1,g}$ sufficiently large. Here, we discuss PB in
these two cases. First, we assume that $C_{1,g}$ is sufficiently large. In
fact, from Eq.~(\ref{C1}), we can see that for the case of strong coupling,
i.e., $g\gg (\kappa ,\gamma )$, if $g^{2}=\Delta _{a}\Delta _{0}$, then $%
C_{1,g}$ is resonantly enhanced and $g^{(2)}(0)\sim \gamma ^{2}/g^{2}\ll 1$.
Here, we have assumed $\kappa =\gamma $ for simplicity. Obviously, this
result is just that obtained in Sec.~\ref{secII-A} based on the energy level
method. More importantly, this behavior appears only when the atomic and
cavity detunings have the same sign.

We now turn to the other case. From Eq.~(\ref{C2}), we can see that if the
signs of the atomic and cavity detunings are opposite, then $C_{2,g}$ can be
minimized by a suitable choice parameters, leading to $g^{(2)}(0)\ll 1$.
Specifically, if we let $C_{2,g}$ vanish, we can obtain very strong photon
antibunching, which is just the PB induced by quantum interference that was
discussed in Sec.~\ref{secII-B}.

Here, we discuss the weak- and strong-coupling regimes. First, if the
coupling is weak, i.e., $g\sim (\kappa ,\gamma )$, then all terms in $\kappa
$ and $\gamma $ must be kept to ensure that $C_{2,g}$ is small enough to
fulfill $g^{(2)}(0)\ll 1$. Hence, the optimal parameter relation should be
the same as Eq.~(\ref{optimalC}). Second, we consider the case of strong
coupling, $g\gg $ $(\kappa ,\gamma )$. Under this condition, the
second-order correlation function can be reduced to
\begin{equation}
\begin{split}
g^{(2)}(0) \simeq{}&\frac{( g^{2}-\Delta _{a}\Delta _{0}) ^{2}}{ 4\Delta
_{0}^{4}(\Delta _{a}^{2}+\Delta _{0}\Delta _{a}-g^{2})^{2}} \\[3pt]
&\times \big\{ 4( g^{2}+\Delta _{a}\Delta _{0}+\Delta _{0}^{2}) ^{2} \\
&\hspace{16pt}+[ \Delta _{a}\gamma +( k+2\gamma ) \Delta _{0}] ^{2}\big\}.
\label{g}
\end{split}%
\end{equation}
It can be seen that when $g^{2}=-\Delta _{0}( \Delta _{a}+\Delta _{0}) $,
Eq.~(\ref{g}) will take the minimal value
\begin{equation}
g^{(2)}(0)\sim \frac{\gamma ^{2}}{g^{2}}\ll 1.  \label{gQI}
\end{equation}%
This shows that even when the optimal parameter condition is not exactly
satisfied, there still exists strong antibunching when the coupling is
sufficiently strong. Of course, when the photons from the system do not
satisfy the optimal parameter condition exactly, the antibunching is weaker
than when they do. This feature makes the PB induced by quantum interference
more flexible and tunable experimentally. Additionally, from the optimal
parameter relation $g^{2}=-\Delta _{0}( \Delta _{a}+\Delta _{0}) $, it can
be seen that if $\vert \Delta _{a} \vert >2g$, then there are two local
minima of $g^{(2)}(0)<1$ located at an atomic detuning $\Delta _{0}=(
-\Delta _{a}\pm \sqrt{\Delta _{a}^{2}-4g^{2}}) /2$; otherwise, if $\vert
\Delta _{a} \vert <2g$, there is no value of $g^{(2)}(0)<1$, which implies
that photons are bunched in the system.

\section{Quantum statistics of cavity photons}

\label{secIII}

In this section, we will present numerical results for $g^{( 2) }(0) $ to
illustrate the statistical properties of the cavity photons. In the
numerical calculations, we employ the master equation for the density matrix
$\rho $ of the system:
\begin{equation}
\frac{\partial \rho }{\partial t}=-i[ H_{I},\rho ] +\frac{\kappa }{2}L[ a]
\rho +\frac{\gamma }{2}L[ \sigma _{-}] \rho ,  \label{MasterEq}
\end{equation}%
where $L[ o] \rho =2o\rho o^{\dagger }-o^{\dagger }o\rho -\rho o^{\dagger }o$
denotes the Lindblad terms accounting for losses to the environment. The
equal-time second-order correlation function can then be evaluated as $g^{(
2) }( 0) =\mathrm{Tr}(\rho a^{\dagger }a^{\dagger }aa) /\mathrm{Tr}(\rho
a^{\dagger }a) ^{2}$ when the system is in a steady state. In the following
numerical calculation, we assume $\kappa =\gamma $ and the largest photon
number of Fock space is restricted to 5, which is sufficient to guarantee the
precision of the numerical results for a weak driving $\epsilon=0.01\gamma$.

\subsection{Fixed coupling strength}

\begin{figure}[tbp]
\centering \includegraphics[width=1\columnwidth]{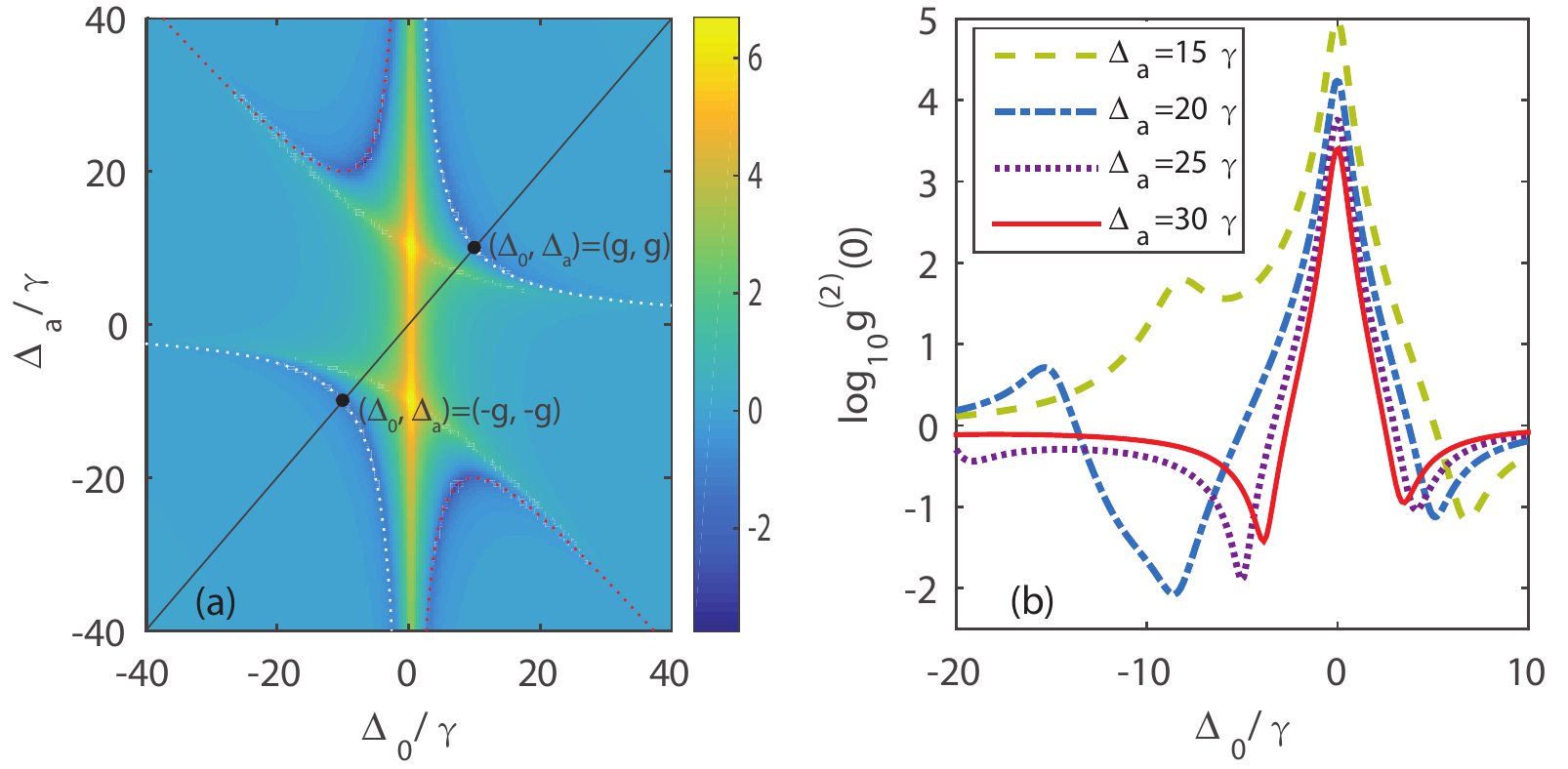}
\caption{(Color online) (a) Second-order correlation function $\log
_{10}g^{( 2) }( 0) $ as a function of the atomic detuning $\Delta _{0}$ and
cavity detuning $\Delta _{a}$. (b) Cross section of (a) along $\Delta _{0}$
with atomic detuning $\Delta _{a}=15\protect\gamma$, $20\protect\gamma$, $25%
\protect\gamma$, and $30\protect\gamma $. The other parameters are taken as $%
\protect\kappa =\protect\gamma $, $g=10\protect\gamma$, and $\protect%
\varepsilon =0.01\protect\gamma $. }
\label{fig2}
\end{figure}

We consider first the case when the system is in the strong-coupling regime
(we take $g=10\gamma $ as an example in the following discussion). We
present contour plots of $g^{(2)}(0)$ as a function of the atomic detuning $%
\Delta _{0}$ and cavity detuning $\Delta _{a}$ in Fig.~\ref{fig2}(a). It can
be seen that when $\Delta _{0}=\Delta _{a}$ (shown by the thin black solid
line), there are two dips ($g^{(2)}(0)\ll 1$) located at $(\Delta
_{0},\Delta _{a})=\pm (10\gamma ,10\gamma )$, corresponding to the
conventional results. However, when the constraint equating the atomic and
cavity detunings is relaxed, we find that the optimal parameter relations
appear in two different parameter regions. In the first of these, the atomic
and cavity detunings have the same sign, i.e., $\Delta _{0}\Delta _{a}>0$,
and are located in the first and third quadrants of the $(\Delta _{0},\Delta
_{a})$ plane. In this region, the optimal parameter relations are two curves
symmetric with respect to the origin $(0,0)$, which is consistent with the
theoretical result given by Eq.~(\ref{6}) and shown by the white dotted
lines in Fig.~\ref{fig2}(a). Of course, the corresponding photon blockade is
induced by the strong nonlinear interaction between the photons. In the
second region, the atomic and cavity detunings have opposite signs, i.e., $%
\Delta _{0}\Delta _{a}<0$, and are located in the second and fourth
quadrants of the $(\Delta _{0},\Delta _{a})$ plane. There are again two
optimal parameter relations represented by curves symmetric with respect to
the origin $(0,0)$, which is very interesting and is observed for the first
time in this work. The numerical optimal parameter relation again confirms
the analytical relation (\ref{optimalC}), indicated by the red dotted lines
in Fig.~\ref{fig2}(a). It should be noted that the PB in this case arises
mainly from quantum interference.

To see the dependence of $g^{(2)}(0)$ on the atomic detuning $\Delta _{0}$
more clearly, we show cross sections of $g^{( 2) }(0) $ for different blue
detunings of the cavity in Fig.~\ref{fig2}(b). It can be seen that when the
atomic detuning is blue, i.e., $\Delta _{0}\Delta _{a}>0$, there is only one
dip, located at $\Delta _{0}=g^{2}/\Delta _{a}$, in the curve of $g^{( 2)
}(0)$. In another region, where the atom is red-detuned with $\Delta
_{0}\Delta _{a}<0$, the situation becomes complicated and dependent on the
detuning of the cavity. When $\Delta _{a}=15\gamma $, which satisfies the
condition $\vert \Delta _{a} \vert <2g$, all values of $g^{( 2) }(0)$ are
greater than $1$, implying that the photons are bunched, as discussed
earlier. When $\vert \Delta _{a} \vert \geq 2g$, taking $\Delta
_{a}=20\gamma $, $25\gamma$, and $30\gamma $ as examples, we see that there
is at least one local minima of $g^{( 2) }(0)$ that is far smaller than $1$,
corresponding to strong photon antibunching. It is interesting to find that
the minimal values of $g^{(2)}(0)$ in the atomic red-detuning regime are
smaller than those in the atomic blue-detuning regime with the same cavity
detuning and coupling strength. Hence, we can obtain more strongly
antibunched photons when $\Delta _{0}\Delta _{a}<0$. For example, with a
cavity detuning $\Delta _{a}=20\gamma $, the minimal value of $g^{( 2)
}(0)\approx 0.012$ at $\Delta _{0}\simeq -9.3\gamma $ in the atomic
red-detuning regime, while the minimal value of $g^{( 2) }(0)\approx 0.077$
at $\Delta _{0}\simeq 5\gamma $ in the atomic blue-detuning regime. The
physical reason here is that quantum interference between different
transition paths may lead to reduced photon population of the two-photon
excitation level. In fact, similar quantum interference-induced strong
antibunching has also been observed and discussed in Ref.~\cite{11}.

\begin{figure}[tbp]
\centering \includegraphics[width=1\columnwidth]{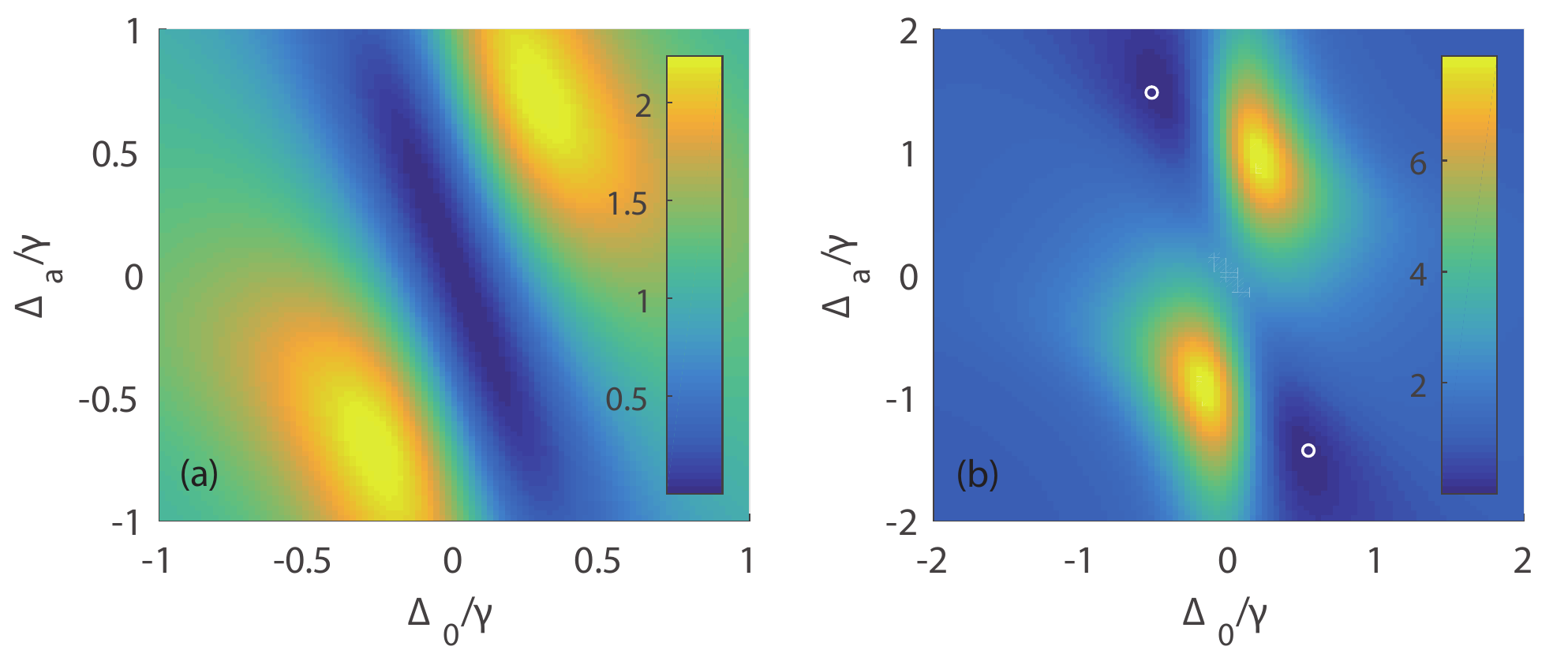}
\caption{(Color online) Second-order correlation function $\log _{10}g^{( 2)
}( 0) $ as a function of the detunings $\Delta _{0}$ and $\Delta_{a}$ with
(a) $g=\protect\gamma /\protect\sqrt{2}$ and (b) $g=\protect\gamma $. The
other parameters are taken as $\protect\kappa =\protect\gamma $ and $\protect%
\varepsilon =0.01\protect\gamma $. }
\label{fig3}
\end{figure}

We now turn to the weak-coupling case. We plot $g^{(2)}(0)$ as a function of
the atomic detuning $\Delta _{0}$ and the cavity detuning $\Delta _{a}$ with
$g=\gamma /\sqrt{2}$ in Fig.~\ref{fig3}(a). When $\Delta _{a}=\Delta _{0}$,
it can be seen that PB appears at just one point, namely, $\Delta
_{a}=\Delta _{0}=0$, which is consistent with the conclusion in Ref.~\cite%
{phonon1}. However, when the constraint equating $\Delta _{0}$ and $\Delta
_{a}$ is relaxed, strong PB appears within the region $(-\gamma /3<\Delta
_{0}<\gamma /3$, $-\gamma <\Delta _{a}<\gamma )$. The optimal relation
between atomic and cavity detunings for strong antibunching is $\Delta
_{a}=-3\Delta _{0}$. We find that the global minimum of $g^{(2)}(0)$ is
located at $( \Delta _{0},\Delta _{a}) =(0,0)$. These features are
consistent with the theoretical prediction from Eq.~(\ref{optimalC}). We
also present a plot of $g^{(2)}(0)$ as a function of the atomic detuning $%
\Delta _{0}$ and the cavity detuning $\Delta _{a}$ with $g=\gamma $ in Fig.~%
\ref{fig3}(b). According to the discussion in Sec.~\ref{secII}, there should
exist two global minima of $g^{(2)}(0)$ at $( \Delta _{0},\Delta _{a}) =\pm
( \gamma /2,-3\gamma /2) $, which is confirmed by the numerical results
indicated by the two white circles in Fig.~\ref{fig3}(b). We can see that
there is no photon antibunching when the signs of the cavity detuning and
the atomic detuning are the same, located in the first and third quadrants
of the $(\Delta _{0},\Delta _{a})$ plane. Hence, we can conclude that the PB
arises mainly from quantum interference in the weak-coupling case, in
contrast to the strong-coupling case.

\subsection{Fixed cavity detuning}

\begin{figure}[tbp]
\centering \includegraphics[width=1\columnwidth]{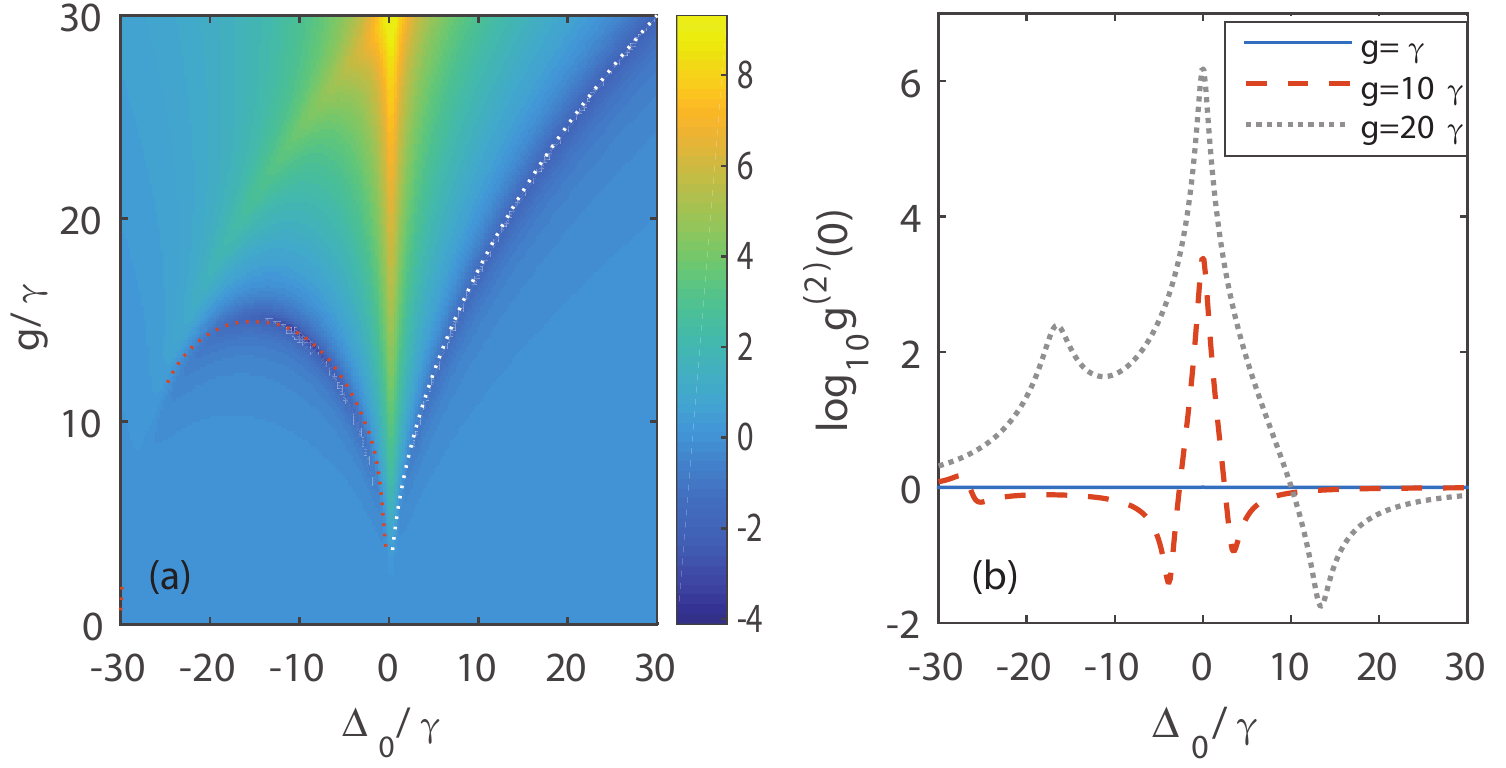}
\caption{(Color online) (a) Second-order correlation function $\log_{10}g^{(
2) }( 0) $ as a function of the detuning $\Delta _{0}$ and coupling strength
$g$. (b) Cross section of $\log _{10}g^{( 2) }( 0) $ along the atomic
detuning $\Delta _{0} $ for $g=\protect\gamma $, $10\protect\gamma$, and $20%
\protect\gamma $. The other parameters are taken as $\protect\kappa =\protect%
\gamma $, $\Delta _{a}=30\protect\gamma $, and $\protect\varepsilon =0.01%
\protect\gamma $. }
\label{fig4}
\end{figure}

In this subsection, we will fix the cavity detuning and consider the
dependence of photon statistical properties on the coupling strength $g$ and
atomic detuning $\Delta _{0}$. Here, we take $\Delta _{a}=30\gamma $ (blue
detuning of the cavity) as an example and plot $g^{(2)}(0)$ as a function of
$\Delta _{0}$ and $g$ in Fig.~\ref{fig4}(a). We can see that $g^{(2)}(0)$
behaves differently in different atomic detuning regimes. In the atomic
blue-detuning regime, PB always appears for suitable choices of atomic
detuning, and the optimal parameter relation curve is parabolic in the
strong-coupling regime. This is confirmed by the white dotted line in Fig.~%
\ref{fig4}(a) and is consistent with the analytical prediction given by Eq.~(%
\ref{6}). The existence of PB is restricted to a finite parameter range in
the atomic red-detuning regime. This can be explained by the fact that if
Eq.~(\ref{optimalC} ) has a real solution for $\Delta _{0}$, then the
coupling strength $g$ must be smaller than $\Delta _{a}/2$. The
corresponding optimal parameter relation is indicated by the red dotted line
in Fig.~\ref{fig4}(a), which also agrees with the analytical result from
Eq.~(\ref{optimalC}). In the weak-coupling regime, the optimal parameter
relation is not close to being satisfied, and hence photon antibunching is
degraded with decreasing $g$.

To reveal the dependence of photon statistics on coupling strength more
clearly, Fig.~\ref{fig4}(b) shows cross sections of $g^{( 2) }(0)$ for
different coupling strengths $g$. It can be seen that $g^{( 2) }(0)=1$ for
all $\Delta _{0}$ when $g=\gamma $. However, for $g=10\gamma $, there are
three local minima of $g^{( 2) }(0)$: one located in the blue-detuning
regime and two in the red-detuning regime. The global minimum of $g^{( 2)
}(0)$ is at $\sim -3.8\gamma $. For stronger coupling, $g=20\gamma $, there
is only one minimum of $g^{( 2) }(0)\ll 1$, located at $\Delta _{0}\simeq
13.3\gamma $.

From the above discussion, it can be seen that the behavior of PB is
strongly dependent on the relation $\Delta _{a}=-( k/\gamma +2) \Delta _{0} $%
, especially in the weak-coupling regime, where the cavity and atomic
detunings have opposite signs. In the following, we discuss further the
relationship between this expression and the PB effect.

\begin{figure}[tbp]
\centering \includegraphics[width=1\columnwidth]{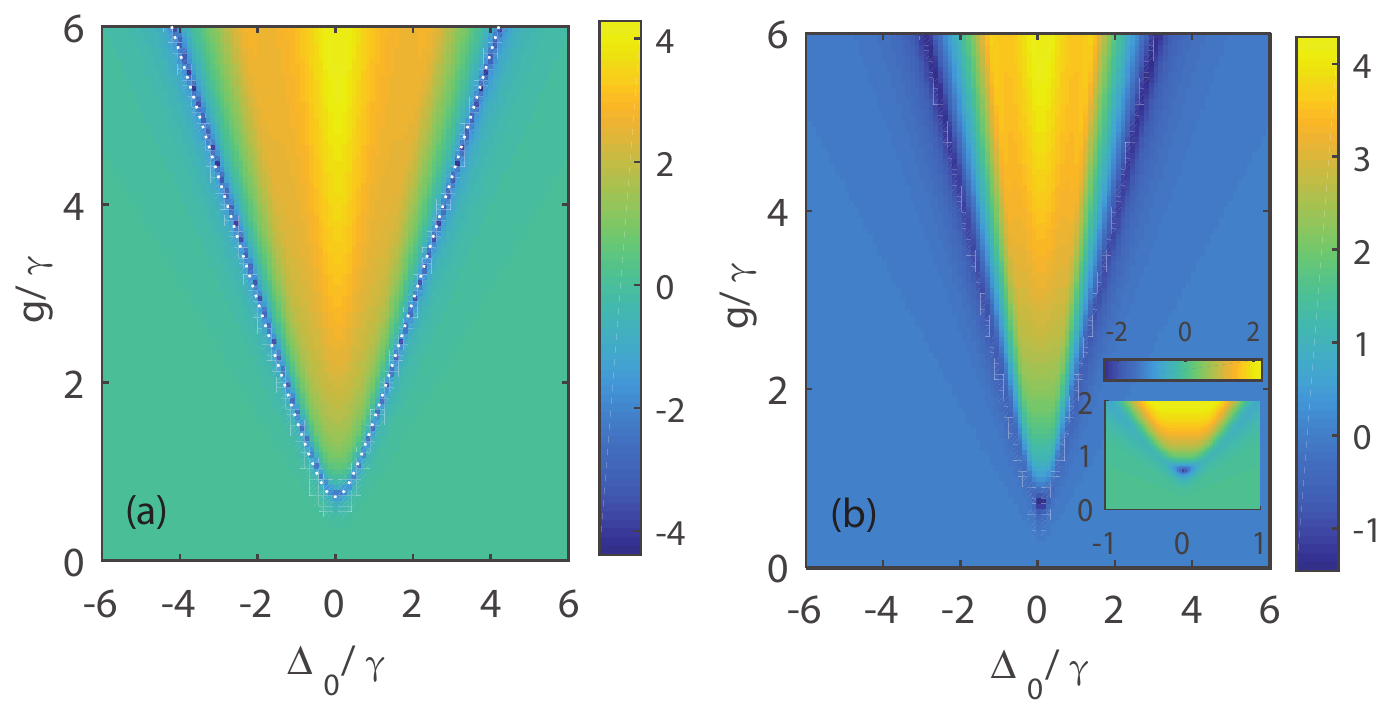}
\caption{(Color online) Second-order correlation function $\log_{10}g^{( 2)
}( 0) $ as a function of the atomic detuning $\Delta _{0}$ and the coupling
strength $g$ for (a) $\Delta _{a}=-3\Delta _{0}$ and (b) $\Delta
_{a}=-5\Delta _{0}$. The other parameters are taken as $\protect\kappa =%
\protect\gamma $ and $\protect\varepsilon =0.01\protect\gamma $. }
\label{fig5}
\end{figure}

Figure~\ref{fig5}(a) shows the dependence of $g^{(2)}(0)$ on the atomic
detuning $\Delta _{0}$ and the coupling strength $g$ when $\Delta _{a}=-( \
k/\gamma +2) \Delta _{0}$. It can be seen that the optimal parameter
relation has the form of a single hyperbola, on which the photons exhibit
strong antibunching. In contrast to the situation discussed above, for all
coupling strengths $g>0.7\gamma $, $g^{(2)}(0)\ll 1$ when $g$ and $\Delta
_{0}$ take their optimal values. This behavior can be interpreted as
follows. According to Eq.~(\ref{optimalC}), if the optimal parameter
condition holds, then both its real and imaginary parts must be zero, i.e., $%
\Delta _{a}=-( k/\gamma +2) \Delta _{0}$ and $g^{2}-2\Delta _{0}^{2}=\gamma
^{2}/2$. In this case, exact quantum destructive interference will result in
very strong antibunching, i.e., $g^{(2)}(0)\ll 1$. From the optimal
relation, we also obtain the lower bound $g_{\min }=\gamma /\sqrt{2}$, which
is consistent with the numerical result.

To more clearly understand the effect of quantum interference on strong
antibunching, we also set $\Delta _{a}=-5\Delta _{0}$ and plot $g^{(2)}(0)$
in Fig.~\ref{fig5}(b). Now the imaginary part of the optimal parameter
relation is nonzero. In this situation, the optimal parameter relation for
exact destructive interference does not hold, which leads to weaker
antibunching. Even at $g\sim \gamma $, there is no antibunching except for $%
g=\gamma /\sqrt{2}$, as shown in Fig.~\ref{fig5}(b).

\section{Conclusion}

\label{secIV}

We have reconsidered PB in an atom--cavity system with two energy levels by
relaxing the constraint that the atomic detuning be equal to the cavity
detuning. We have found that even when this constraint is not satisfied,
there still exists a strong antibunching effect. When the atomic and cavity
detunings have the same sign, PB is induced by nonlinear photon interaction
in the strong-coupling regime. In this situation, the corresponding optimal
parameter relation becomes $g^{2}=\Delta _{a}\Delta _{0}$. More importantly,
the results show that in both the strong- and weak-coupling regimes, there
is strong antibunching due to quantum interference when the atomic and
cavity detunings are of opposite sign The corresponding optimal parameter
relations are $\Delta _{a}=-( k/\gamma +2) \Delta _{0}$ and $g^{2}+\Delta
_{a}\Delta _{0}+\Delta _{0}^{2}=( \gamma +k) \gamma /4$. This new feature
has been observed for the first time in this work. Further analysis shows
that no PB is induced by nonlinear photon interaction in the strong-coupling
regime, in contrast to the case when the atomic and cavity detunings have
the same sign. This results obtained here should allow the construction of
single-photon sources based on PB in two-level atom--cavity systems that
offer greater tunability and flexibility.

\acknowledgments
This project was supported by the National Natural Science Foundation of
China (Nos. 11664014, 11504145 and 11364021).

\end{document}